\documentclass[11pt]{article}

\usepackage{fullpage}

\usepackage{booktabs} 

\usepackage{latexsym}
\usepackage{url}
\usepackage{listings}
\usepackage{xspace}
\usepackage{color}
\usepackage{graphicx}

\usepackage{amsmath}
\usepackage{amssymb}

\newcommand{\add}[1]{\texttt{add(#1)}\xspace}
\newcommand{\rem}[1]{\texttt{rem(#1)}\xspace}
\newcommand{\con}[1]{\texttt{con(#1)}\xspace}
\newcommand{\CAS}[1]{\texttt{CAS(#1)}\xspace}

\title{A more Pragmatic Implementation of the Lock-free, Ordered, Linked List}
\author{Jesper Larsson Tr\"aff,  Manuel P\"oter\\
  TU Wien\\
  Faculty of Informatics\\
  Institute of Computer Engineering, Research Group Parallel Computing\\
  Favoritenstrasse 16, 1040 Vienna, Austria}
\date{October 29th, 2020}

\begin{document}
\maketitle

\begin{abstract}
  The lock-free, ordered, linked list is an important, standard example
  of a concurrent data structure. An obvious, practical drawback of
  textbook implementations is that failed compare-and-swap (\CAS{})
  operations lead to retraversal of the entire list (retries), which
  is particularly harmful for such a linear-time data structure. We
  alleviate this drawback by first observing that failed \CAS{}
  operations under some conditions do not require a full retry, and
  second by maintaining approximate backwards pointers that are used
  to find a closer starting position in the list for operation
  retry. Experiments with both a worst-case deterministic benchmark,
  and a standard, randomized, mixed-operation throughput benchmark on
  three shared-memory systems (Intel Xeon, AMD EPYC, SPARC-T5) show
  practical improvements ranging from significant, to dramatic, several
  orders of magnitude.
\end{abstract}

\section{Introduction}

The lock-free, ordered, singly linked list as proposed
in~\cite{Harris01,Michael02} is a textbook example of a concurrent
data structure~\cite{HerlihyShavit12,Scott13}. The data structure
supports lock-free insertion (\add{}) and deletion (\rem{}), and
wait-free contains (\con{}) operations on items identified by a unique
key. The lock-free implementation is actually quite subtle.  The
ordering condition and a relaxed invariant makes it possible to do
with a single-word compare-and-swap operation (\CAS{}), and all
operations can be shown to be linearizable even though linearization
does not always happen at fixed points in the code (\CAS{}
operations). The lock-free data structure has many direct and indirect
applications, notably in the implementation of concurrent skiplists
and hash
tables~\cite{Michael02,Pugh90,ShalevShavit06,SundellTsigas05}.

An obvious, practical drawback of the standard implementation is the
draconic action on failed \CAS{} operations: A retraversal from the
head of the list is simply initiated.  For a linear-time data
structure, this is particularly harmful since it in the worst case
entails traversal of the whole list. For any individual thread, this
can happen indefinitely, since the implementation is not
starvation-free. Available implementations seem to implement ordered,
lock-free lists in this way~\cite{David15,Gramoli15,Guerraoui16,libcds}.

In~\cite{FomitchevRuppert04}, this problem was addressed by
maintaining a doubly linked list. The implementation is more
complex involving two atomic flags and extra \CAS{} operations; the
paper focuses on a theoretical analysis, and gives no experimental
results, and none seem to have been prominently reported.

The present, short paper gives some most pragmatic improvements to the
textbook implementation that seem to benefit practical performance
from significantly to dramatically. The improvements first seek to
avoid complete retraversal of the list wherever possible by examining
the reason for each failed \CAS{} operation. In both \add{}, \rem{}
and the crucial internal operation for finding the position in the
list on which to add or remove an item, there are such possibilities
(this may well be done in other implementations, but we have not found
any published claims). Second, by extending the list items with a
predecessor pointer that is maintained approximately by conditional,
atomic stores and loads, all failed \CAS{} operations traverse
the list backwards only to a point where a new, forwards find
operation can be started. The backward pointers only have to fulfill
that for any list item, there is a path back to the head (sentinel)
element of the list.  The improvements change the instances where the
list operations linearize, but we claim that the implementation
remains linearizable largely as the textbook implementation.

Unfortunately, the introduction of backward pointers and cursor significantly
complicate the problem of safe memory reclamation. The implementation
benchmarked here does only simple memory reclamation after each experiment.
How to do proper memory reclamation with these improvements is at the moment
beyond this paper.

\section{Improvement and Implementation}

The standard lock-free, ordered linked list implementation maintains
the invariants that items are in key order and that an item is in the
list \emph{iff} it is reachable from the head and not marked. The mark
(flag) kept with each list item is a stolen bit in the item's pointer
to the next larger item, such that mark and pointer can be checked and
updated atomically by a single \CAS{} operation. Textbook
implementations use a non-visible search function for locating the
position where a given key could be; this function has the additional,
crucial task of linking out items that have become marked. A \CAS{}
operation is used for linking out a marked item, for inserting a new
item, and for marking an item for deletion. Each of these \CAS{}
operations check both that the item from which the operation is
performed has not become marked, and that the pointer to the next
larger item has not changed.

We observe the following.
\begin{itemize}
\item
  Upon failure of the \CAS{} operation in the search function, if the
  item on which the operation failed has \emph{not} become marked
  (only the pointer changed due to another thread having had success in
  linking out the item), there is no reason to restart the search from
  the head of the list; it suffices to reread the next pointer of the
  item.
\item
  In the \rem{} operation, if the \CAS{} operation fails due to the
  item having become marked, the delete can be linearized as failed,
  since the item has been removed by another thread. The linearization
  point is not the failed \CAS{}, though, but the earlier of this and
  the point just before an overlapping \add{} operation is about to
  successfully insert an item with the same key. If instead the \CAS{}
  failed due to the pointer to the next item having changed, the
  thread can simply try again to mark the item. This can be repeated
  until the item to be deleted has become marked, successfully by some
  thread. Once the node is marked it is logically deleted and will be
  removed eventually. So a failure of the subsequent \CAS{} to unlink
  the node from the list can simply be ignored.
\item
  In an \add{} operation, if the \CAS{} operation fails due to the
  next pointer having changed (but not the mark), the search for the
  position where to add can be continued from the item, there is no
  reason to go back to the head of the list. All that is needed is to
  reread the pointer.
\end{itemize}

\begin{lstlisting}[float=t,caption={The search function that links out and ``physically removes'' a marked item, textbook implementation and mild improvements. \texttt{LOAD} and \texttt{CAS} abbreviate the C atomic operations.},label=lst:pos]
void pos(long key, list_t *list) {
  node_t *pred, *succ, *curr, *next;

retry:
#ifdef TEXTBOOK
  pred = list->head;
#else
  pred = list->pred;
  if (ismarked(LOAD(&pred->next)) || key <= pred->key)
    pred = list->head;
#endif

  curr = getpointer(LOAD(&pred->next));

  do {
    succ = LOAD(&curr->next);
    while (ismarked(succ)) {
      succ = getpointer(succ);
      if (!CAS(&pred->next, &curr, succ)) {
#ifdef TEXTBOOK
        goto retry;
#else
        next = LOAD(&pred->next);
        if (ismarked(next)) goto retry;
        succ = next;
#endif
      }

      curr = getpointer(succ);
      succ = LOAD(&succ->next);
    }

    if (key <= curr->key) {
      list->pred = pred;
      list->curr = curr;
      return;
    }
    pred = curr;
    curr = getpointer(LOAD(&curr->next));
  } while (1);
}
\end{lstlisting}

\begin{lstlisting}[float=t,caption={The \rem{} operation, textbook implementation and mild improvement. Also the possible improvement by using an atomic
    fetch-and-or operation (macro \texttt{FAO}) instead of the \texttt{CAS}-loop is shown.},label=lst:rem]
int rem(long key, list_t *list) {
  node_t *pred, *succ, *node;
  node_t *markedsucc;

  do {
    pos(key, list);
    pred = list->pred;
    node = list->curr;
    if (node->key != key) return 0; // not there

#ifdef TEXTBOOK
    succ = getpointer(LOAD(&node->next)); // unmarked
    markedsucc = setmark(succ);

    if (!CAS(&node->next, &succ, markedsucc))
      continue;
#else
#ifdef FETCH
    succ = FAO(&node->next,MARK_BIT);
    if (ismarked(succ)) return 0;
#else    
    succ = LOAD(&node->next);
    do {
      if (ismarked(succ)) return 0;
      markedsucc = setmark(succ);
      if (CAS(&node->next, &succ, markedsucc)) break;
    } while (1);
#endif
#endif
    
    CAS(&pred->next, &node, succ);
    return 1;
  } while (1);
}
\end{lstlisting}

The three observations lead to three mild, pragmatic improvements of
the textbook implementation. The first (search function) and second
(\rem{}) are illustrated with concrete code snippets from our
implementation (that is available from the authors) in Listing~\ref{lst:pos}
and Listing~\ref{lst:rem}. The third mild improvement (in \add{}) is
handled by the search function which checks the mark of the current
element.  Implementation is done in C11 using the standard atomic
operations with the C11 memory model (acquire-release); \texttt{LOAD},
\texttt{STORE} and \texttt{CAS} are macros for these. It is worth
noting that these changes do not introduce any issues regarding memory
reclamation; any of the commonly used techniques will do.

The changes to the \rem{} operation makes another (platform specific)
improvement possible, namely to use an atomic fetch-and-or operation
to set the delete mark on the next pointer. On architectures that
natively support this operation this would have the advantage that the
marking operation cannot fail. We also benchmarked this potential
improvement in the course of this paper by relying on the
corresponding C11 operation. We note that x86 architectures can support
and atomic (lock) or operation, but not an atomic fetch-and-or.

The more intrusive improvement which actually entails the milder
improvements for free, extends list items with a backwards pointer to
a previous, smaller key element. On a \CAS{} failure, the search
function will simply have the task to go backwards in the list, that
is through smaller key items, until an unmarked element is found from
which the search in increasing key order can be started.

\begin{lstlisting}[float=t,caption={The complete search operation with backward pointers. With this search function, \add{} and \rem{} implementations can be kept as they are (textbook). \texttt{LOAD}, \texttt{STORE} and \texttt{CAS} abbreviate the C atomic operations.},label=lst:newpos]
void pos(long key, list_t *list) {
  node_t *pred, *succ, *curr, *next;

  pred = list->pred;

retry:
  while(ismarked(LOAD(&pred->next)) || key<=pred->key)
    pred = LOAD(&pred->prev);

  curr = getpointer(LOAD(&pred->next));
  do {
    succ = LOAD(&curr->next);
    while (ismarked(succ)) {
      succ = getpointer(succ);
      if (!CAS(&pred->next, &curr, succ)) {
        next = LOAD(&pred->next);
        if (ismarked(next)) goto retry;
        succ = next;
      } else STORE(&succ->prev, pred);

      curr = getpointer(succ);
      succ = LOAD(&succ->next);
    }

    if (LOAD(&curr->prev) != pred)
      STORE(&curr->prev, pred);

    if (key <= curr->key) {
      list->pred = pred;
      list->curr = curr;
      return;
    }
    pred = curr;
    curr = getpointer(LOAD(&curr->next));
  } while (1);
}
\end{lstlisting}

The search function with backward pointer is shown in
Listing~\ref{lst:newpos}.  In order for the implementation to be
correct, the backward pointers only have to fulfill that for any item,
there is a path via backward pointers back to the head of the
list. When a new item is inserted into the list, the backwards pointer
of the successor is set with an atomic store to point to the new
element. Likewise, when an item is removed, the backwards pointer of
the successor item is updated to skip the removed item. This suffices
to maintain the invariant, but through long sequences of (concurrent)
insertions and deletions, backwards pointers can become imprecise in
that they skip many items that are actually in the list. As seen in
Listing~\ref{lst:newpos}, we try to maintain more precise backwards
pointers by updating them during forwards traversals. Since updates
with atomic stores are expensive due to cache coherence activity, we
only update a pointer if a test (with non-atomic, relaxed loads) shows
that a pointer is not correct. Also, when a marked item is linked out,
the backwards pointer of its successor is updated to skip the linked
out item. One reason for this is that memory reclamation can (only) be
done when there are no pointers to a removed item.

In contrast to the algorithm of~\cite{FomitchevRuppert04}, this
optimistic implementation has no extra flag to be maintained with
possibly expensive (failing) \CAS{} operations.

The final, highly pragmatic improvement is to exploit throughout the
doubly linked structure of the list. Each thread maintains a cursor
item in the list. On its next operation, depending on the key of the
item to be located, the find operation will search either forwards
(increasing key order) or backwards (smaller key order) in the list
from the cursor. Each \add{}, \rem{} and \con{} operations sets the
cursor to the item before the located item.

Obviously, keeping a cursor to an item in the middle of the list
reduces the average runtime complexity for a singly linked
structure as well, since we only have to search from head to cursor,
or from cursor to end, depending on the key of the item to be located.

Unfortunately, the introduction of backward pointers significantly
complicates the problem of safe memory reclamation, because it can happen
that a backwards pointer references a node that has already been removed
from the list. In order to retire a node for reclamation, it has to
be ensured that it is not referenced by any next or backwards pointer.
Proper memory reclamation with these improvements is currently outside
the scope this work.

\section{Experimental results}

We have conducted a number of experiments with the described C11
implementations on different platforms to illustrate concrete,
pragmatic benefits of the observations and improvements we
discussed. Experiments have been done on three standard multi-core
systems:
\begin{enumerate}
\item
  A 64-core AMD EPYC system with two 32-core AMD EPYC 7551 sockets at 2.9GHz
\item
  An 80-core Intel Xeon with 8 E7-8850 sockets at 1.06GHz
\item
  A 64-core SPARC v9 system with eight 8-core SPARC-T5 sockets at
  3.6GHz and 8x SMT.
\end{enumerate}

We have used two different benchmarks.
\begin{itemize}
\item
Deterministic worst-case benchmark: Starting from an empty list, each
tread performs the following three sequences each of length of $n$:
\con{$k(i)$}, \add{$k(i)$}, \con{$k(i)$}, \add{$k(i)$} for $i=0$ to
$n-1$, then \con{$k(i)$}, \rem{$k(i)$}, \con{$k(i)$}, \rem{$k(i)$}
from $i=n-1$ to $i=0$, finally \con{$k(i)$} for $i=0$ to $n-1$. The
key function $k(i)$ is chosen either such that each thread $t$ has its
own sequence of keys, disjoint from all other threads, $k(i)=t+ip$
($p$ number of threads), or such that all threads have the same key
sequences, $k(i)=i$. Due to the linear search in the ordered lists,
the sequential behavior per thread is $O(pn^2)$ (for disjoint keys) or
$O(n^2)$ (for same keys) steps.
\item
Standard random operation mix benchmark: Keys are chosen uniformly at
random in an interval $[0,U-1]$ (\texttt{random\_r}). The list is
prefilled with some number of items $f$, then a mix of randomly chosen
\add{}, \rem{} and \con{} operations is performed with predefined
probability for each operation (here we report only for the mix
$10-10-80$). Each thread performs the same number of operations $c$, and
threads have different seeds for the random number generator (we use
the thread-safe \texttt{random\_r() generator}, except on SPARC since
this function is not available on Solaris). For chosen $f$ and $U$
the number of elements of the list will not vary too much.
\end{itemize}

We use OpenMP to manage threads and time the benchmarks. The
benchmarks can also be configured such that each thread operates on a
private list, such that there is no interaction required between
threads. In this configuration, we can use either the lock-free
implementation, or a standard, sequential (doubly or singly linked)
list implementation. These configurations can give an idea of the
system and memory overheads when there is no actual interaction
between threads. We do not report on the thread private behavior here.
We report on six implementation variants, namely a) \emph{draconic}
(textbook implementation), b) \emph{singly} linked list with mild
improvements, c) \emph{doubly} linked list with approximate backward
pointers and retry from from head of list, d) \emph{singly-cursor}
singly linked list with per thread retry from the last recorded
position (cursor) in the list, e) \emph{singly-fetch-or} singly linked
list with per thread retry from the last recorded position (cursor)
and atomic fetch-and-or in the \rem{} operations, and f)
\emph{doubly-cursor} doubly linked list with per thread retry from the
last recorded position (cursor) in the list.  On Intel and AMD we used
the standard \texttt{glibc} memory allocator; on SPARC we used the
\texttt{libumem} allocator. All implementations and code used in the
experiments are available from the authors.

Results for fixed number of threads on the different platforms for the
two benchmarks are shown in Tables~\ref{tab:detbenchmarkEPYC1},
\ref{tab:detbenchmarkEPYC2}, \ref{tab:ranbenchmarkEPYC},
\ref{tab:detbenchmarkXeon1}, \ref{tab:detbenchmarkXeon2},
and~\ref{tab:ranbenchmarkXeon}. We report throughput over the number
of operations, and also count the number of failed \CAS{} operations,
the number of retries, the total number of list item traversals in the
search operation, and the total number of traversals in contains
operations, and the total number of successful \add{} and \rem{}
operations. The results in the tables are for one single specific run
of the benchmarks. Runs differ, but in a tolerable range.

\begin{figure*}[!tb]
	\centering
	\includegraphics[width=\textwidth]{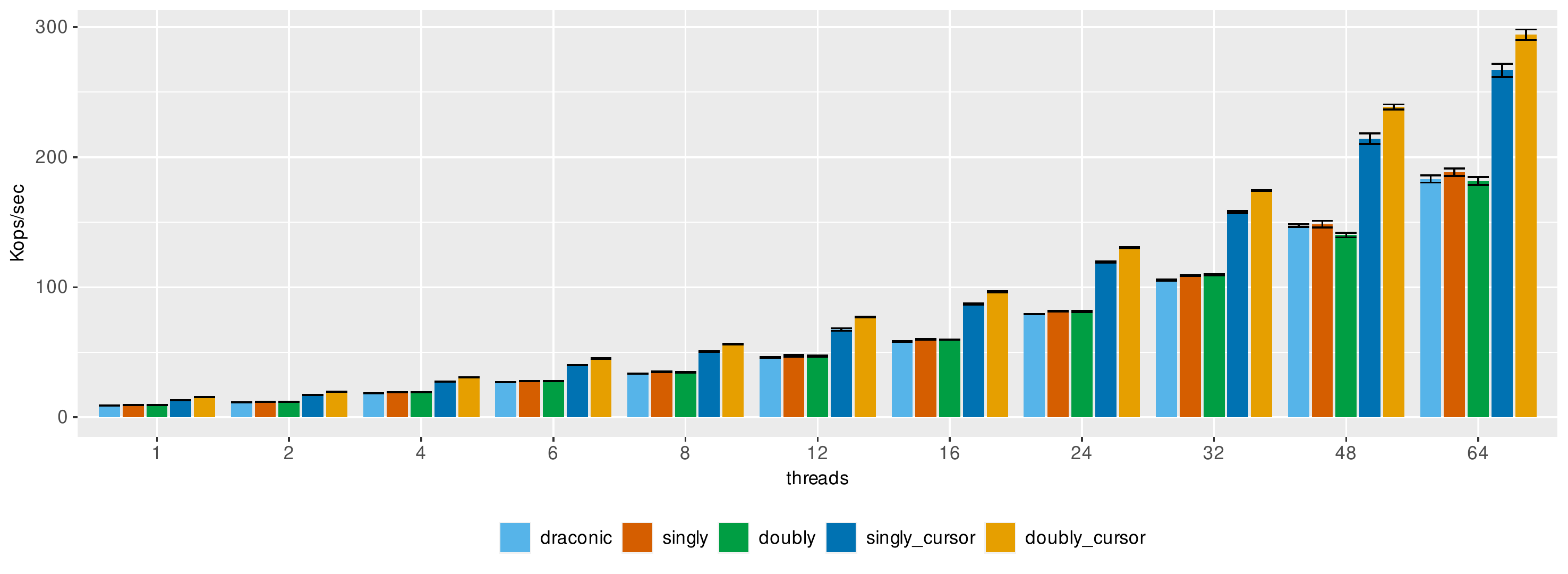}
	\caption{Scalability with threads of random operation mix
          benchmark, AMD EPYC system, $c = 50000, f = 16384, U =
          32768$. Operation Mix 25\% \add{}, 25\% \rem{}, 50\% \con{}}
	\label{fig:threads-AMD}
\end{figure*}

\begin{figure*}[!tb]
	\centering
        \includegraphics[width=\textwidth]{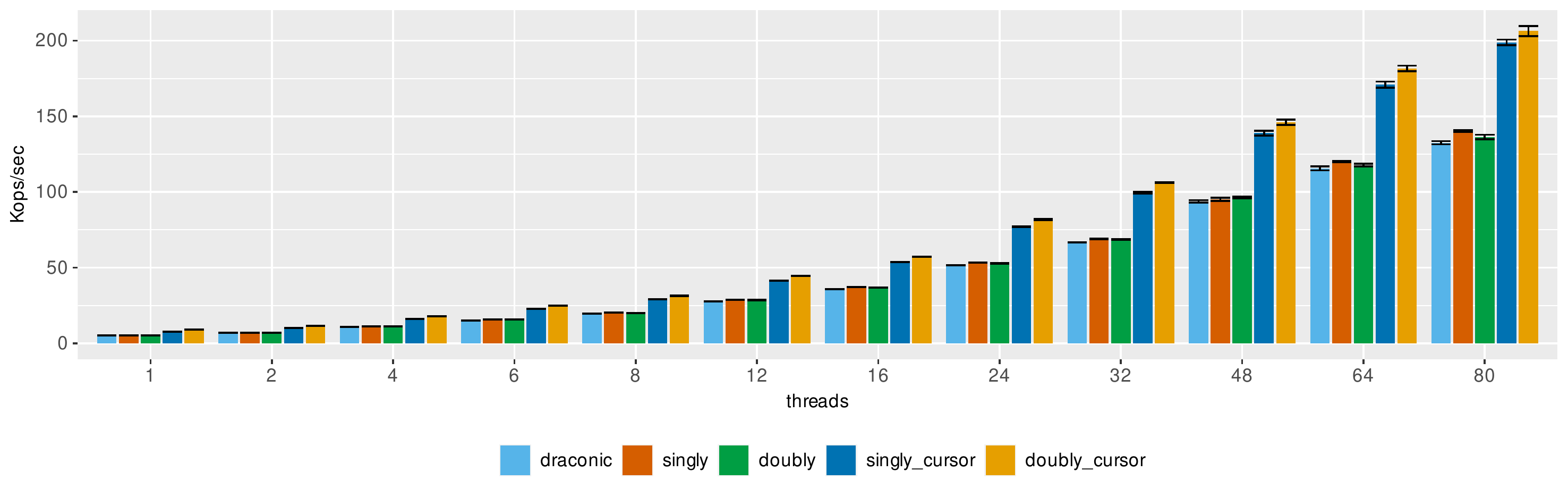}
	\caption{Scalability with threads of random operation mix
          benchmark, Intel Xeon system, $c = 50000, f = 16384, U =
          32768$. Operation Mix 25\% \add{}, 25\% \rem{}, 50\% \con{}}
	\label{fig:threads-Intel}
\end{figure*}

\begin{figure*}[!tb]
	\centering
        	\includegraphics[width=\textwidth]{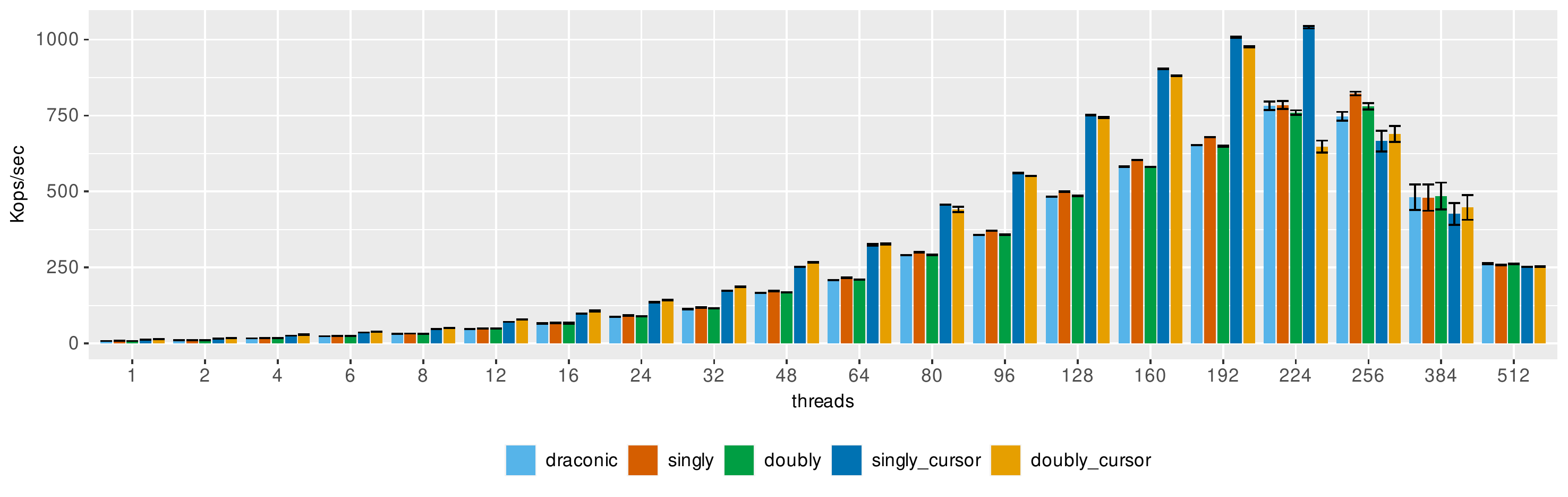}

	\caption{Scalability with threads of random operation mix
          benchmark, SPARC-T5 system, $c = 50000, f = 16384, U =
          32768$. Operation Mix 25\% \add{}, 25\% \rem{}, 50\% \con{}}
	\label{fig:threads-Sparc}
\end{figure*}

We investigate the (weak) scalability of five of the six (excluding the
atomic fetch-and-or variant) variants with the
random operation mix benchmark. Here, we plot the mean throughput of
$5$ experiments. The scalability results are shown in
Figures~\ref{fig:threads-AMD}, \ref{fig:threads-Intel}
and~\ref{fig:threads-Sparc}.  The experiments were run with a key
range of 32768 and an update ratio of 50\% (25\% \add{}, 25\% \rem{});
the lists were prefilled with 16384 items.  These settings are
comparable to those used in \cite{Gramoli15}.

In the tables and plots, $p$ denotes the number of started threads,
$n$ the list length for the deterministic benchmark, $c$ the number of
operations per thread for the random mix benchmark (thus weak scaling
in the scalability experiments, since the number of operations per
thread is kept fixed for increasing $p$), $f$ the number of prefilled
elements, and $U$ the upper bound for the key range.

\begin{table*}[t]
  \caption{Deterministic benchmark $k(i)=i$, AMD EPYC system, $p=64,
    n=100000$. Operation breakdown: ``adds'' is the number of successful \add{} operations, ``rems'' the number of successful \rem{} operations, ``cons'' the number of element traversals over all \con{} operations, ``trav'' the number of list element traversals in the search function, ``fail'' the number of \CAS{} failures, and ``rtry'' the number of retries in the search function. Variants: a) draconic, b) singly, c) doubly, d) singly-cursor, e) singly-fetch-or, f) doubly-cursor.}
\label{tab:detbenchmarkEPYC1}
\begin{scriptsize}
\begin{tabular}{crrrrrrrrr}
  \toprule
  Variant &
  Time  & Total ops & Throughput & adds & rems & cons & trav & fail & rtry \\
  & (ms) &  & (Kops/s) & & & & & &  \\
  \midrule
  a) &
162841.78 & 57600000 & 353.72 & 143909 & 143909 & 1218424056670 & 1218522498969 & 2715 & 2534 \\
b) &
145519.31 & 57600000 & 395.82 & 145091 & 145091 & 1182002909974 & 921060012934 & 4697 & 64 \\
  c) &
141036.26 & 57600000 & 408.41 & 129492 & 129492 & 1271917666836 & 850820221201 & 18113 & 0 \\
  d) &
85334.28 & 57600000 & 674.99 & 133575 & 133575 & 38267280 & 1236958075255 & 19376 & 22 \\
  e) &
78687.72 & 57600000 & 732.01 & 123834 & 123834 & 38260212 & 1236723136990 & 16235 & 24 \\
  f) &
481.28 & 57600000 & 119681.20 & 100333 & 100333 & 76141143 & 56187324 & 591036 & 16450 \\
  \bottomrule
  \end{tabular}
\end{scriptsize}
\end{table*}


\begin{table*}
  \caption{Deterministic benchmark $k(i)=t+ip$, AMD EPYC system,
    $p=64, n=10000$.
    Operation breakdown: ``adds'' is the number of successful \add{} operations, ``rems'' the number of successful \rem{} operations, ``cons'' the number of element traversals over all \con{} operations, ``trav'' the number of list element traversals in the search function, ``fail'' the number of \CAS{} failures, and ``rtry'' the number of retries in the search function. Variants: a) draconic, b) singly, c) doubly, d) singly-cursor, e) singly-fetch-or, f) doubly-cursor.}
\label{tab:detbenchmarkEPYC2}
\begin{scriptsize}
\begin{tabular}{crrrrrrrrr}
  \toprule
  Variant &
  Time  & Total ops & Throughput & adds & rems & cons & trav & fail & rtry \\
  & (ms) &  & (Kops/s) & & & & & &  \\
  \midrule
a) &
860636.43 & 5760000 & 6.69 & 640000 & 640000 & 808043265652 & 823284224246 & 46720 & 22816 \\
  b)  & 
787290.12 & 5760000 & 7.32 & 640000 & 640000 & 808320969602 & 608143267099 & 38738 & 0 \\
  c) & 
807260.63 & 5760000 & 7.14 & 640000 & 640000 & 801938148075 & 602327832241 & 28103 & 0 \\
  d) &
504290.83 & 5760000 & 11.42 & 640000 & 640000 & 201364518449 & 804041576627 & 45914 & 0 \\
  e) &
578657.21 & 5760000 & 9.95 & 640000 & 640000 & 202104065277 & 807669700081 & 51486 & 1 \\
  f) & 
285.97 & 5760000 & 20142.13 & 640000 & 640000 & 58322194 & 5202640 & 17221 & 25 \\
  \bottomrule
\end{tabular}
\end{scriptsize}
\end{table*}

\begin{table*}
\caption{Random operation mix benchmark, AMD EPYC system, $p=64,
  c=1000000, f=1000, U=10000$. Operation Mix 10\% \add{}, 10\% \rem{},
  80 \% \con{}. Operation breakdown: ``adds'' is the number of successful \add{} operations, ``rems'' the number of successful \rem{} operations, ``cons'' the number of element traversals over all \con{} operations, ``trav'' the number of list element traversals in the search function, ``fail'' the number of \CAS{} failures, and ``rtry'' the number of retries in the search function. Variants: a) draconic, b) singly, c) doubly, d) singly-cursor, e) singly-fetch-or, f) doubly-cursor.}
\label{tab:ranbenchmarkEPYC}
\begin{scriptsize}
\begin{tabular}{crrrrrrrrr}
  \toprule
  Variant &
  Time  & Total ops & Throughput & adds & rems & cons & trav & fail & rtry \\
  & (ms) &  & (Kops/s) & & & & & &  \\
  \midrule
  a) &
42633.85 & 64000000 & 1501.15 & 3198087 & 3203070 & 127961354252 & 32034345892 & 23399 & 23189 \\
  b) & 
44169.92 & 64000000 & 1448.95 & 3197813 & 3202696 & 127951211375 & 26665833669 & 18441 & 1 \\
  c) & 
43643.95 & 64000000 & 1466.41 & 3197408 & 3202406 & 128000172533 & 26651455978 & 20517 & 0 \\
  d) & 
28527.89 & 64000000 & 2243.42 & 3197537 & 3202414 & 85313098938 & 21336289626 & 25488 & 0 \\
e) &
30076.08 & 64000000 & 2127.94 & 3196895 & 3201757 & 85299471964 & 21332876996 & 24305 & 0 \\
f) & 
20028.06 & 64000000 & 3195.52 & 3201780 & 3199113 & 85221592358 & 21296756185 & 15544 & 0 \\
  \bottomrule
  \end{tabular}
\end{scriptsize}
\end{table*}

\begin{table*}[t]
\caption{Deterministic benchmark $k(i)=i$, Intel Xeon system, $p=80,
  n=100000$. Operation breakdown: ``adds'' is the number of successful \add{} operations, ``rems'' the number of successful \rem{} operations, ``cons'' the number of element traversals over all \con{} operations, ``trav'' the number of list element traversals in the search function, ``fail'' the number of \CAS{} failures, and ``rtry'' the number of retries in the search function. Variants: a) draconic, b) singly, c) doubly, d) singly-cursor, e) singly-fetch-or, f) doubly-cursor.}
\label{tab:detbenchmarkXeon1}
\begin{scriptsize}
\begin{tabular}{crrrrrrrrr}
  \toprule
  Variant &
  Time  & Total ops & Throughput & adds & rems & cons & trav & fail & rtry \\
  & (ms) &  & (Kops/s) & & & & & &  \\
  \midrule
  a) &
314154.80 & 72000000 & 229.19 & 111618 & 111618 & 1574094573004 & 1574104025889 & 3655 & 301 \\
  b) &
264868.53 & 72000000 & 271.83 & 111229 & 111229 & 1553877268888 & 1188479836763 & 457 & 6 \\
  c) &
265895.34 & 72000000 & 270.78 & 108673 & 108673 & 1599794312194 & 1151226363104 & 2795 & 1 \\
  d) &
184202.03 & 72000000 & 390.88 & 121969 & 121969 & 47879590 & 1560938517400 & 8405 & 1 \\
  e) &
195536.43 & 72000000 & 368.22 & 126514 & 126514 & 47874986 & 1555697504918 & 6189 & 50 \\
  f) &
3366.28 & 72000000 & 21388.60 & 100189 & 100189 & 95827925 & 69781787 & 57781 & 138 \\
  \bottomrule
  \end{tabular}
\end{scriptsize}
\end{table*}

\begin{table*}
  \caption{Deterministic benchmark $k(i)=t+ip$, Intel Xeon system,
    $p=80, n=10000$. Operation breakdown: ``adds'' is the number of successful \add{} operations, ``rems'' the number of successful \rem{} operations, ``cons'' the number of element traversals over all \con{} operations, ``trav'' the number of list element traversals in the search function, ``fail'' the number of \CAS{} failures, and ``rtry'' the number of retries in the search function. Variants: a) draconic, b) singly, c) doubly, d) singly-cursor, e) singly-fetch-or, f) doubly-cursor.}
\label{tab:detbenchmarkXeon2}
\begin{scriptsize}
\begin{tabular}{crrrrrrrrr}
  \toprule
  Variant &
  Time  & Total ops & Throughput & adds & rems & cons & trav & fail & rtry \\
  & (ms) &  & (Kops/s) & & & & & &  \\
  \midrule
  a) & 
3482460.95 & 7200000 & 2.07 & 800000 & 800000 & 1254565652071 & 1446526400768 & 375424 & 114672 \\
  b) & 
3176148.60 & 7200000 & 2.27 & 800000 & 800000 & 1276788904826 & 957852619867 & 287865 & 1 \\
  c) & 
2917187.37 & 7200000 & 2.47 & 800000 & 800000 & 1276828251706 & 957639298722 & 254600 & 0 \\
  d) &
2147669.87 & 7200000 & 3.35 & 800000 & 800000 & 319401544377 & 1277563045294 & 488450 & 69 \\
  e) &
2129745.54 & 7200000 & 3.38 & 800000 & 800000 & 319037405095 & 1276676865674 & 484925 & 78 \\
  f) & 
584.10 & 7200000 & 12326.70 & 800000 & 800000 & 69175184 & 6697424 & 33751 & 439 \\
  \bottomrule
  \end{tabular}
\end{scriptsize}
\end{table*}

\begin{table*}
\caption{Random operation mix benchmark, Intel Xeon system, $p=80,
  c=1000000, f=1000, U=10000$. Operation Mix 10\% \add{}, 10\% \rem{},
  80 \% \con{}. Operation breakdown: ``adds'' is the number of successful \add{} operations, ``rems'' the number of successful \rem{} operations, ``cons'' the number of element traversals over all \con{} operations, ``trav'' the number of list element traversals in the search function, ``fail'' the number of \CAS{} failures, and ``rtry'' the number of retries in the search function. Variants: a) draconic, b) singly, c) doubly, d) singly-cursor, e) singly-fetch-or, f) doubly-cursor.}
\label{tab:ranbenchmarkXeon}
\begin{scriptsize}
\begin{tabular}{crrrrrrrrr}
  \toprule
  Variant &
  Time  & Total ops & Throughput & adds & rems & cons & trav & fail & rtry \\
  & (ms) &  & (Kops/s) & & & & & &  \\
  \midrule
  a) & 
79940.23 & 80000000 & 1000.75 & 3997685 & 4002627 & 159819683981 & 40004806528 & 29670 & 24868 \\
  b) & 
78877.44 & 80000000 & 1014.23 & 3997774 & 4002741 & 159771141405 & 33291359815 & 24771 & 1 \\
  c) & 
78152.47 & 80000000 & 1023.64 & 3997991 & 4002934 & 159882542042 & 33279948727 & 25699 & 0 \\
  d) &
56889.93 & 80000000 & 1406.22 & 3994914 & 3999857 & 106554989656 & 26647195005 & 28602 & 2 \\
  e) &
56785.59 & 80000000 & 1408.81 & 3996829 & 4001795 & 106534595059 & 26642102852 & 28528 & 1 \\
  f) & 
43498.16 & 80000000 & 1839.16 & 3996761 & 4001799 & 106589641350 & 26636431208 & 16761 & 1 \\
  \bottomrule
  \end{tabular}
\end{scriptsize}
\end{table*}

\begin{table*}
  \caption{Deterministic benchmark $k(i)=i$, SPARC-T5 system, $p=64,
    n=100000$. Operation breakdown: ``adds'' is the number of successful \add{} operations, ``rems'' the number of successful \rem{} operations, ``cons'' the number of element traversals over all \con{} operations, ``trav'' the number of list element traversals in the search function, ``fail'' the number of \CAS{} failures, and ``rtry'' the number of retries in the search function. Variants: a) draconic, b) singly, c) doubly, d) singly-cursor, f) doubly-cursor.}
\label{tab:detbenchmarkSparc1}
\begin{scriptsize}
\begin{tabular}{crrrrrrrrr}
  \toprule
  Variant &
  Time  & Total ops & Throughput & adds & rems & cons & trav & fail & rtry \\
  & (ms) &  & (Kops/s) & & & & & &  \\
  \midrule
  a) & 
  514603.24 & 57600000 & 111.93 & 113010 & 113010 & 1264254674628 & 1264624767976 & 8356 & 7233 \\
  b) & 
  434627.41 & 57600000 & 132.53 & 108395 & 108395 & 1275648845411 & 955962198414 & 61709 & 26 \\
  c) & 
  407405.79 & 57600000 & 141.38 & 129056 & 129056 & 1273179327218 & 858421075762 & 168084 & 0 \\
  d) & 
  286471.04 & 57600000 & 201.07 & 108319 & 108319 & 38292797 & 1272875360528 & 103677 & 165 \\
  f) &
  173.51 & 57600000 & 331962.89 & 169502 & 169502 & 76406418 & 55414912 & 719558 & 4338 \\
  \bottomrule
  \end{tabular}
\end{scriptsize}
\end{table*}

\begin{table*}
  \caption{Deterministic benchmark $k(i)=t+ip$, SPARC-T5 system,
    $p=64, n=10000$. Operation breakdown: ``adds'' is the number of successful \add{} operations, ``rems'' the number of successful \rem{} operations, ``cons'' the number of element traversals over all \con{} operations, ``trav'' the number of list element traversals in the search function, ``fail'' the number of \CAS{} failures, and ``rtry'' the number of retries in the search function. Variants: a) draconic, b) singly, c) doubly, d) singly-cursor, f) doubly-cursor.}
\label{tab:detbenchmarkSparc2}
\begin{scriptsize}
\begin{tabular}{crrrrrrrrr}
  \toprule
  Variant &
  Time  & Total ops & Throughput & adds & rems & cons & trav & fail & rtry \\
  & (ms) &  & (Kops/s) & & & & & &  \\
  \midrule
  a) &
  4207815.61 & 5760000 & 1.37 & 640000 & 640000 & 818179538925 & 818586650089 & 4196 & 3926 \\
  b) & 3686586.39 & 5760000 & 1.56 & 640000 & 640000 & 818018742600 & 613614553897 & 33093 & 12 \\
  c) & 3687992.32 & 5760000 & 1.56 & 640000 & 640000 & 818285400314 & 613707337468 & 319493 & 59 \\
  d) & 2661259.96 & 5760000 & 2.16 & 640000 & 640000 & 204614565262 & 818219910239 & 425616 & 42 \\
  f) & 968.42 & 5760000 & 5947.82 & 640000 & 640000 & 47807228 & 5779948 & 288786 & 62911 \\
  \bottomrule
\end{tabular}
\end{scriptsize}
\end{table*}

\begin{table*}
\caption{Random operation mix benchmark, SPARC-T5 system, $p=64,
  c=1000000, f=1000, U=10000$. Operation Mix 10\% \add{}, 10\% \rem{},
  80 \% \con{}. Operation breakdown: ``adds'' is the number of successful \add{} operations, ``rems'' the number of successful \rem{} operations, ``cons'' the number of element traversals over all \con{} operations, ``trav'' the number of list element traversals in the search function, ``fail'' the number of \CAS{} failures, and ``rtry'' the number of retries in the search function. Variants: a) draconic, b) singly, c) doubly, d) singly-cursor, f) doubly-cursor.}
\label{tab:ranbenchmarkSparc}
\begin{scriptsize}
\begin{tabular}{crrrrrrrrr}
  \toprule
  Variant &
  Time  & Total ops & Throughput & adds & rems & cons & trav & fail & rtry \\
  & (ms) &  & (Kops/s) & & & & & &  \\
  \midrule
  a) & 
40806.24 & 64000000 & 1568.39 & 3200918 & 3205988 & 120278045492 & 30185936395 & 48740 & 48521 \\
  b) & 
39319.52 & 64000000 & 1627.69 & 3203939 & 3208845 & 120295394002 & 25351649568 & 42377 & 0 \\
  c) & 
40012.58 & 64000000 & 1599.50 & 3200439 & 3205376 & 120329461579 & 25310904222 & 38018 & 0 \\
  d) &
27749.98 & 64000000 & 2306.31 & 3200810 & 3205729 & 81939214206 & 20509159845 & 58033 & 2 \\
  f) & 
27130.05 & 64000000 & 2359.01 & 3199878 & 3204457 & 87100467429 & 21785440781 & 27313 & 0 \\
  \bottomrule
  \end{tabular}
\end{scriptsize}
\end{table*}


\section{Discussion}

The mild improvements over the draconic textbook implementation
clearly reduce the number of retries; as a consequence it also reduces
the completion time and improves the throughput, but not as much as
expected. The (emulated) atomic fetch-and-or operation as expected brings no
improvement over the corresponding improved singly linked list with cursor.

On average, the speedup in the scalability experiments is
about 3.4\%. Making the list doubly linked induces additional
overhead that actually costs performance.  In the scalability experiments,
the average speedup of the doubly linked list in comparison to draconic is
only about 1.9\%.

However, the situation changes when we keep a cursor to the last used
item.  In case of the doubly linked list this allows us to starting
search in the backwards/forwards direction from last item position;
for the singly linked list it still cuts the list into two pieces,
allowing us to start searching either from the last position or from
head, depending on the key of the item to be located. This can improve
performance dramatically, in particular for the doubly linked list in
the deterministic benchmark (orders of magnitude).

But also in the random mix benchmark the cursor based implementations
are significantly faster. When using a cursor, the ability to search
backwards in doubly linked list pays off, resulting in significantly
better performance, at least on Intel and AMD; on SPARC, the cursor
based singly/doubly linked list implementations perform more or less
on par.

The mild improvements (with cursor) are easy, unintrusive improvements
to the standard, textbook implementation of the lock-free, ordered
linked list with significant enough performance improvements to be
considered, also for more complex algorithms (skip lists and hash
tables) that build on the linked list data structure. These
improvements do not comprise the chosen memory reclamation scheme. The
approximate backward pointers in the doubly linked improvements can be
extremely beneficial in certain cases (the deterministic benchmarks
that were designed to highlight this), but come at cost, and
complicate memory reclamation. A possible application is a
simplification in the implementation of the Stamp-It memory
reclamation system~\cite{Traff18:reclamation,Traff18:stampit}.

\bibliographystyle{plain}
\bibliography{lockfreelist} 

\end{document}